\documentclass[twocolumn,showpacs,preprintnumbers,amsmath,amssymb]{revtex4}

\usepackage{graphicx}
\usepackage{dcolumn}
\usepackage{bm}
\usepackage{amsmath,amssymb}

\begin{document}

\preprint{APS/123-QED}

\title{Mott transition in one-dimensional boson-fermion mixtures}

\author{Yousuke Takeuchi}
\affiliation{%
Department of Quantum Matter, ADSM, Hiroshima University,\\ 1-3-1 Kagamiyama, 
Higashi-Hiroshima 739-8530, JAPAN
}%

\author{Hiroyuki Mori}
\affiliation{
Department of Physics, Tokyo Metropolitan University,\\
Minamiohsawa 1-1, Hachioji-shi, Tokyo 192-0397, JAPAN\\
}%

\date{\today} 

\begin{abstract}
We numerically investigated Mott transitions and mixing-demixing transitions in one-dimensional boson-fermion mixtures at a commensurate filling. The mixing-demixing transition occurred in a qualitatively similar manner to incommensurate filling cases. We also found the Mott insulator phase appeared in both the mixing and the demixing states as the fermion-boson interaction or the boson-boson interaction increased. Phase diagrams were obtained in interaction parameter space.
\end{abstract}

\pacs{Valid PACS appear here}
\maketitle
\section{\label{sec1}Introduction}
Ultra-cold alkaline atom gases have opened up new possibilities for experimentalists and theorists. The trapped atomic gases allow tuning of physical parameters and provide various kinds of systems from weak to strong coupling regimes. For example, commensurate filling bosonic systems can be constructed and Mott transition has been one of attractive topics studied theoretically \cite{jaksch}-\cite{Ho} and experimentally \cite{greiner}. 

Easy construction of low-dimensional systems is another advantage of the trapped atoms. In particular one-dimensional boson-fermion mixtures have attracted attention in terms of mixing-demixing transition, and theoretical studies have been reported \cite{Das}-\cite{RotandBu}. In Ref. \cite{we} we carried out Monte Carlo simulations to investigate the mixing-demixing transition in a system at an incommensurate filling of fermions and boson, and found that the system undergoes the transition from the mixing to demixing phases as the boson-fermion interactions increase. Figure 1 is a schematic phase diagram in the $U_{fb}-U_{bb}$ plane where $U_{fb}$ is the fermion-boson interaction and $U_{bb}$ the boson-boson interaction. The solid curve presents the phase boundary between the mixing and demixing phases. An exact solution assures the dashed straight line, $U_{fb}=U_{bb}$, is in the mixing phase \cite{AdiandEug}.
\begin{figure}
\resizebox{45mm}{!}{\includegraphics{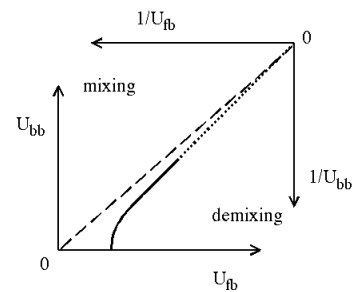}}
\caption{\label{fig1}Schematic diagram of mixing and demixing phases of incommensurate boson-fermion mixtures. An exact solution assures that the dashed line ($U_{bb}=U_{fb}$)is in the mixing phase.} 
\end{figure}

In this paper we examined the Mott transitions and the mixing-demixing transitions in systems at a commensurate filling on a one-dimensional periodic lattice. We know single-component commensurate systems become Mott insulators as inter-particle interactions increase. In the two-component systems, however, the situation is rather complicated partly because the definition of the term "commensurability" has some variation. We could consider a case where the total filling is commensurate but the filling of each component is incommensurate (Case 1), or a case where the total filling as well as the filling of each component are all commensurate (Case 2). Roth and Burnett numerically studied a mixture of 4 fermions and 4 bosons on an 8-site lattice, i.e. the Case 2, and showed the presence of mixed Mott insulator and demixed Mott insulator at large fermion-boson interactions and large boson-boson interactions \cite{RotandBu}. We performed Monte Carlo simulations to investigate the Mott transition and the mixing-demixing transition in the Case 1 and 2 and see how the phase diagram in Fig. 1 changes in the commensurate-filling cases.
\section{\label{sec2}Model}
We consider a mixture with $N_f$ spinless fermions and $N_b$ bosons on an $N$-site 
lattice. To let the system undergo the Mott transition, we fixed the total number of particles $N_{tot}=N_f+N_b$ equal to $N$. We employed Bose-Fermi Hubbard Hamiltonian to describe the interacting particles:
 \begin{eqnarray}
\mathcal{H}&=&-\sum_i\sum_{\alpha=f,b}\left[ t_{\alpha}(a_{\alpha,i}^\dagger 
a_{\alpha,i+1}+h.c.)+\mu_{\alpha} n_{\alpha,i} \right] \nonumber\\
& &{}+\sum_i \left[ \frac{U_{bb}}{2} n_{b,i}(n_{b,i}-1)+U_{fb}n_{f,i}n_{b,i} \right],
\end{eqnarray}
where $a_{\alpha,i}^\dagger$ and $a_{\alpha,i}$ are respectively creation and annihilation 
operators for fermions ($\alpha =f$) or bosons ($\alpha =b$) on the $i$-th 
site, and $n_{\alpha,i}=a_{\alpha,i}^\dagger a_{\alpha,i}$.
Hopping energy and chemical potential are denoted respectively by $t_{\alpha}$ and $\mu_{\alpha}$.
$U_{bb}$ and $U_{fb}$ are on-site boson-boson and fermion-boson repulsive 
interactions respectively. We set $t_f=t_b=1$ as an energy unit.

We performed Monte Carlo simulations with the world line algorithm to measure the system behavior \cite{Hirsch, BatandSca}. Temperature was fixed to $T=1/15$ and the Trotter decomposition number $L$ to 100. The number of the sites $N$ was set to 14 and the total filling to 1, i.e.  $N_{tot}=N=14$. In the simulation of the Case 1 we chose $N_f=10$ and $N_b=4$, and in the simulation of the Case 2 we used $N_f=N_b=7$.

To observe the mixing-demixing transition, we measured a correlation function \cite{we},
\begin{equation}
C =\langle (n_{b,i-1}+n_{b,i+1})n_{b,i}\rangle - \langle 
(n_{f,i-1}+n_{f,i}+n_{f,i+1})n_{b,i}\rangle , \label{m1}
\end{equation}
which tends to be smaller, or negatively larger, in mixing states than in demixing states.

To observe the Mott transition, we measured a current-current correlation function $J_{tot}$ in the zero-frequency limit \cite{BatandSca, Batetal},
\begin{equation}
J_{tot}=\lim_{\omega\to 0} \langle j_{tot}(\omega )j_{tot} (-\omega )\rangle ,
\end{equation}
where $j_{tot}$ presents the total current density of fermions and bosons. $J_{tot}$ becomes zero when the system is in the Mott insulator phase.
\section{Result}
\subsection{Case 1}
At first let us see some limiting cases. In the limit of $U_{bb}\rightarrow\infty$, i.e. in a hardcore boson case, the mixing-demixing transition was not observed and the system always stayed in the mixing state. This can be understood in the following way. The fermion (boson) gains the kinetic energy by hopping to a boson- (fermion-) occupied site, and the gain should be larger in the mixing state than in the demixing state, because in the mixing state the fermions (bosons) have a lot of chances to hop to the site occupied by the boson (fermion). In the limit of $U_{fb}\rightarrow\infty$ with a finite $U_{bb}$, on the other hand, the system was always in the demixing state. 

What about the Mott transition in these cases? Figure 2 shows the behavior of the correlation function $J_{tot}$ with infinite boson-boson interactions (a) and that with infinite fermion-boson interactions (b).
\begin{figure}
\begin{tabular}{cc}
\resizebox{42mm}{!}{\includegraphics{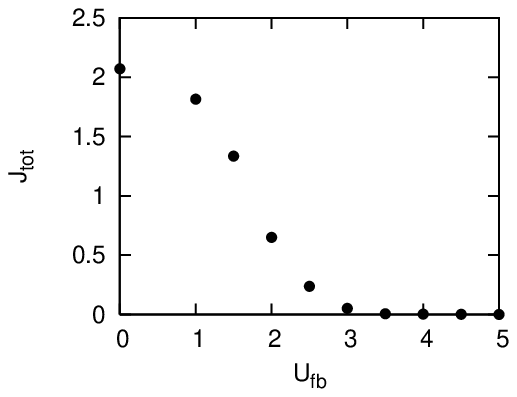}}&
\resizebox{42mm}{!}{\includegraphics{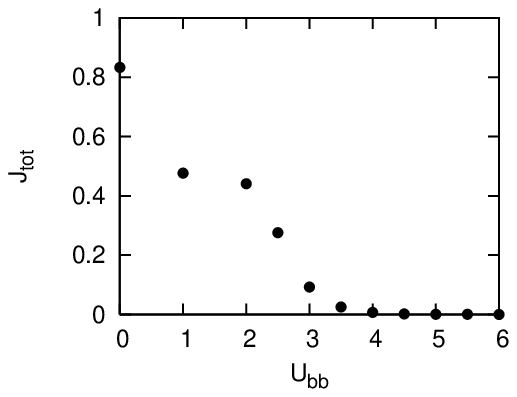}}\\
(a)&(b) 
\end{tabular}
\caption{\label{fig2}The stiffness $J_{tot}$ as a function of $U_{fb}$ with infinite $U_{bb}$ (a) and as a function of $U_{bb}$ with infinite $U_{fb}$ (b).}
\end{figure}
As we can see in the figure, the system undergoes the Mott transition at $U_{fb}\sim 2$ (a), and at $U_{bb}\sim 2.5$ (b).

Next we studied another limit, $U_{bb}=0$. The system did not show the Mott transition in this case, because the bosons can easily overlap each other creating empty sites, which makes the system soft. However the system undergoes the mixing-demixing transition. Figure 3 shows the correlation function $C$ as a function of $U_{fb}$. The uprising curve of the correlation function indicates a transition from the mixing to the demixing states at about $U_{fb}=1.5$. The demixing state can be seen in the snapshot of Fig. 3(b) where the fermions and the bosons stay away from each other. 
\begin{figure}
\begin{tabular}{c}
 \resizebox{45mm}{!}{\includegraphics{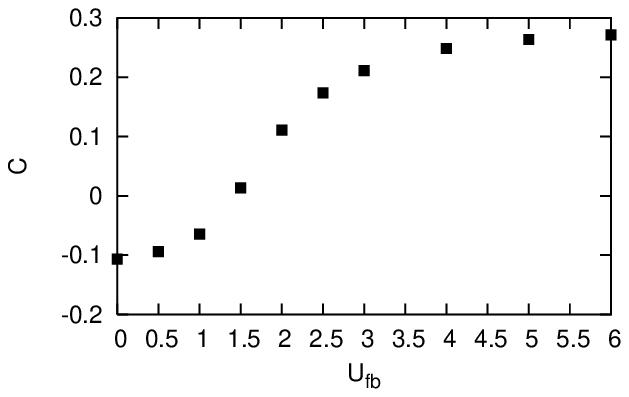}}\\
(a)\\
\end{tabular}
\begin{tabular}{ll}
\resizebox{30mm}{!}{\includegraphics{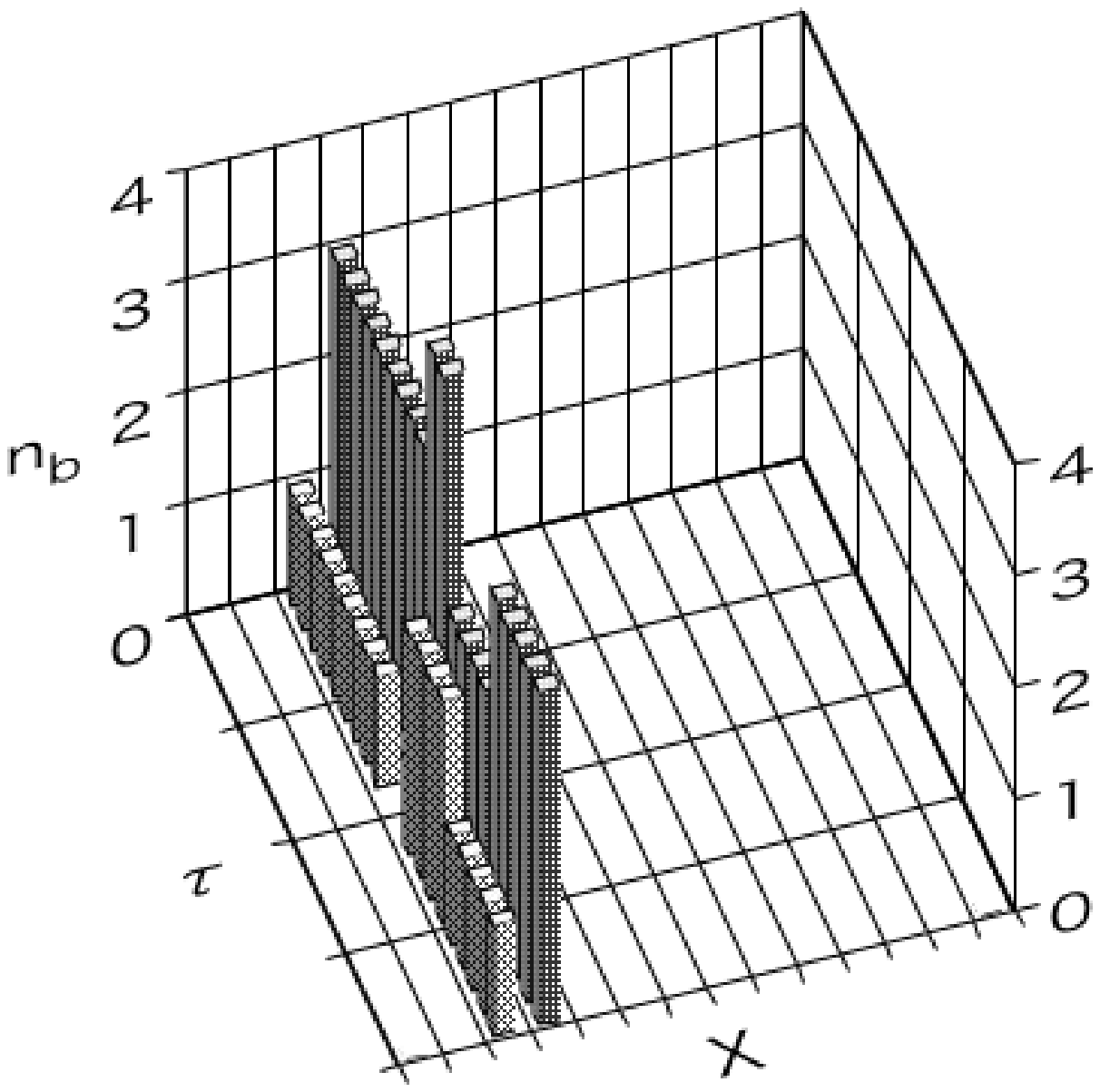}}&
\resizebox{30mm}{!}{\includegraphics{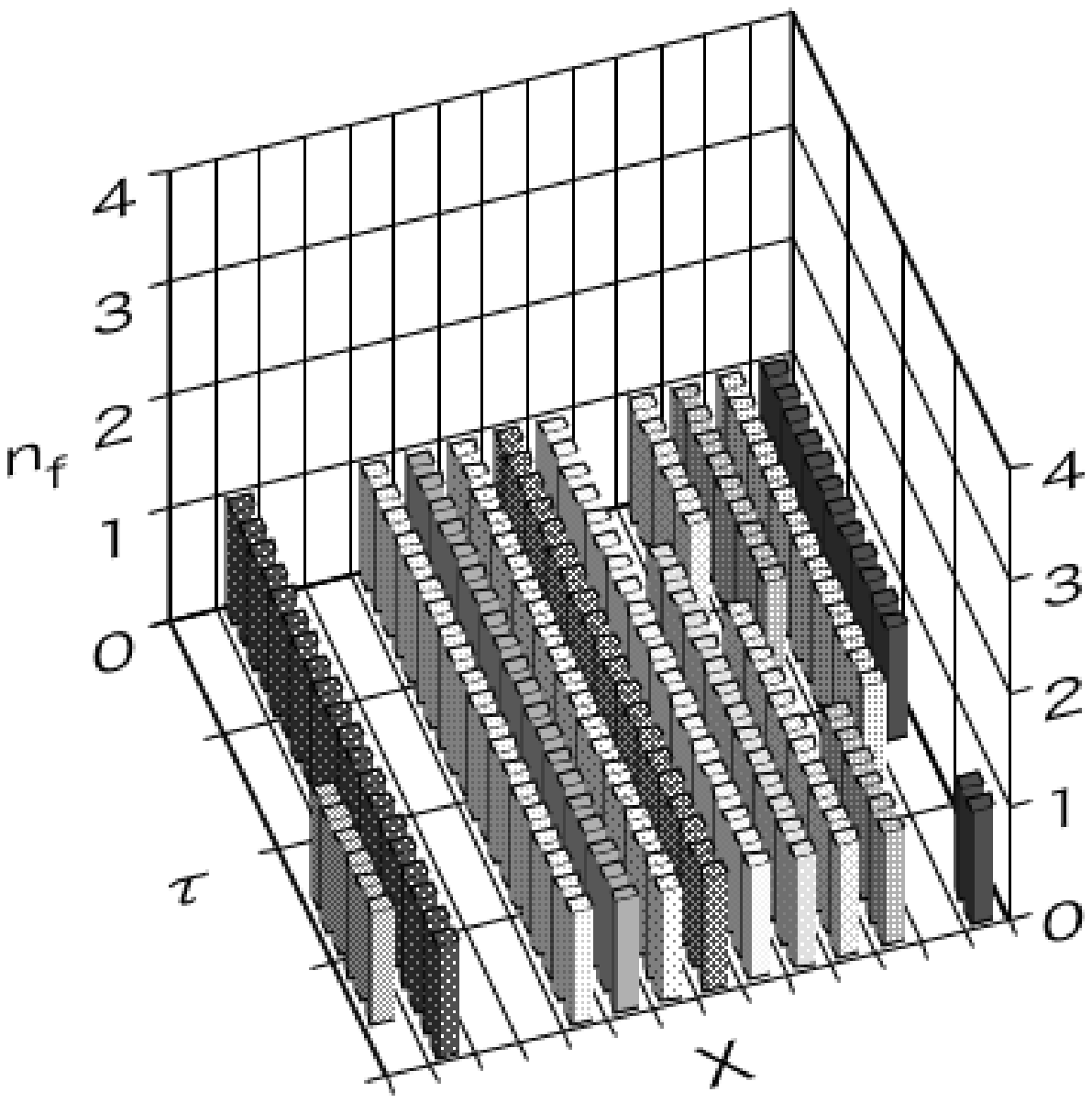}}
\end{tabular}\\
(b)
\caption{\label{fig3} (a) The correlation function $C$ as a function of $U_{fb}$ with $U_{bb}=0$. (b) A snapshot of boson configuration (left) and fermion configuration (right) at $U_{fb}=5$ and $U_{bb}=0$. $X$ is the real space axis and $\tau$ is the imaginary time axis. $n_{b(f)}$ shows the number of the bosons (fermions) on each site.}
\end{figure}

In the above limits, the system showed only either of the Mott transition or the mixing-demixing transition. Now we study the behavior of the system in the parameter region with finite $U_{fb}$ and finite $U_{bb}$. In Fig. 4 we show the correlation functions $C$ and $J$, fixing $U_{fb}=5$ (a) and $U_{fb}=14$ (b). The behavior of $C$ and $J$ indicates the occurrence of the mixing-demixing transition and of the Mott transition both in (a) and (b). The qualitative difference between (a) and (b) is that the mixing-demixing transition occurs almost simultaneously with the Mott transition in (a), while the system undergoes the Mott transition first and then the mixing-demixing transition in (b).
\begin{figure}
\begin{tabular}{cc}
\resizebox{42mm}{!}{\includegraphics{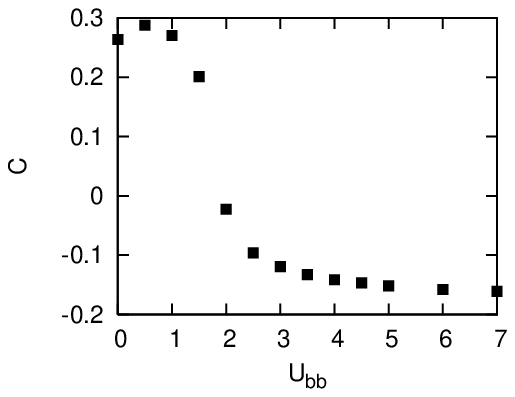}} & 
\resizebox{42mm}{!}{\includegraphics{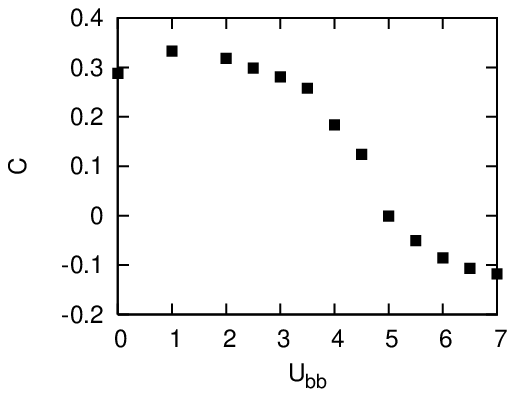}}\\
\resizebox{42mm}{!}{\includegraphics{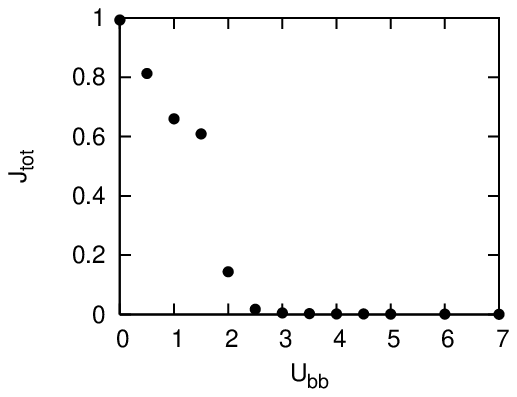}} & 
\resizebox{42mm}{!}{\includegraphics{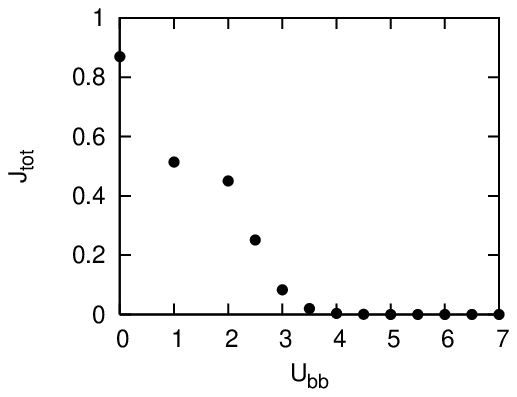}} \\
(a) & (b)
\end{tabular}
\caption{\label{fig4} The correlation functions $C$ and $J_{tot}$ as a function of $U_{bb}$ with $U_{fb}=5$ (a) and $U_{fb}=14$ (b).}
\end{figure}
\subsection{Case 2}
In the Case 2 we have $N_f=N_b=7$ on the 14-site lattice. As in the Case 1 we measured the correlation functions $C$ and $J$. In the limit of $U_{bb}\rightarrow\infty$ the system is in the mixing phase and undergoes the Mott transition at around $U_{fb}=2.5$, and in the limit of $U_{fb}\rightarrow\infty$ the system is in the demixing phase and undergoes the Mott transition at around $U_{bb}=3.5$. When $U_{bb}=0$ the system changes from the mixing state to the demixing state at around $U_{fb}=1$ and never exhibits the Mott transition.

Figure 5 shows $C$ and $J$ in the intermediate parameter region. Their behaviors are quite similar to those in Fig. 4 of the Case 1.
\begin{figure}
\begin{tabular}{cc}
\resizebox{42mm}{!}{\includegraphics{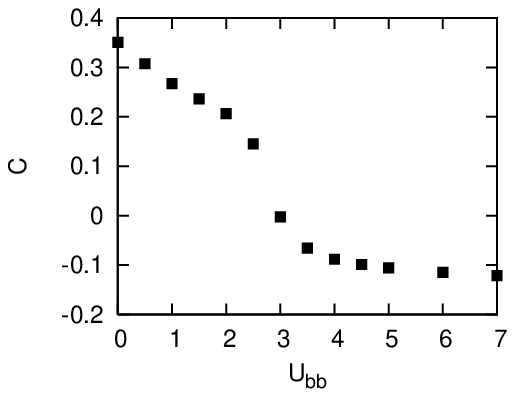}} & 
\resizebox{42mm}{!}{\includegraphics{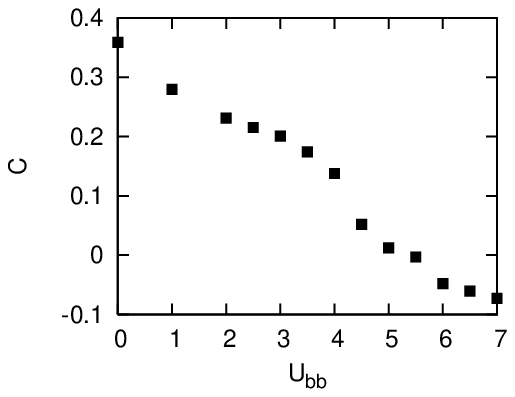}}\\
\resizebox{42mm}{!}{\includegraphics{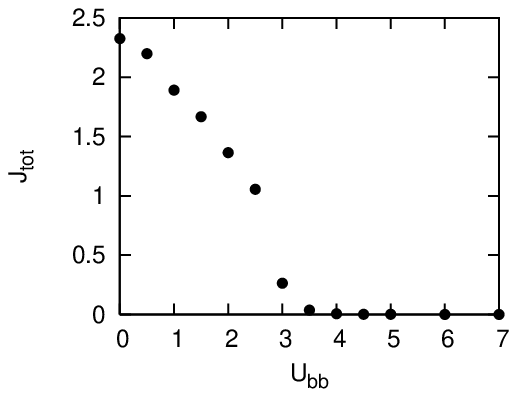}} & 
\resizebox{42mm}{!}{\includegraphics{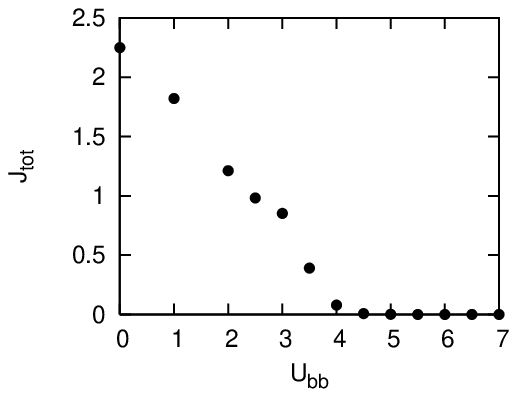}} \\
(a) & (b)
\end{tabular}
\caption{\label{fig5} The correlation functions $C$ and $J_{tot}$ as a function of $U_{bb}$ with $U_{fb}=7$ (a) and $U_{fb}=14$ (b).}
\end{figure}
\subsection{Phase diagram}
Figure 6 (a) and (b) are the phase diagrams of the Case 1 and 2 respectively, produced in the $U_{fb}-U_{bb}$ plane. It is noted that the transition line between the mixing and demixing phases is qualitatively equivalent to the one in the case of incommensurate filling as in Fig. 1. The phase diagrams are also consistent with the result in Ref. \cite{AdiandEug} which proved that the $U_{fb}=U_{bb}$ line stays in the mixing phase.

We do not find any essential difference between the two diagrams (a) and (b), indicating that the phase diagram is determined by the commensurability of the total filling and not very sensitive to whether the filling of each component is commensurate or not.
\begin{figure}
\begin{tabular}{c}
\resizebox{55mm}{!}{\includegraphics{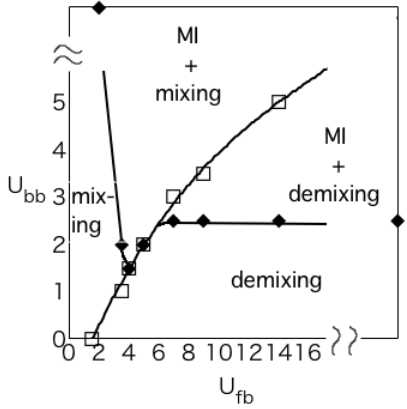}}\\(a)\\ 
\resizebox{55mm}{!}{\includegraphics{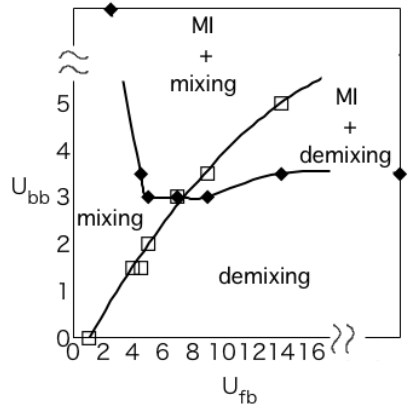}}\\(b)
\end{tabular}
\caption{\label{fig6} Phase diagrams of the Case 1 (a) and Case 2 (b). $U_{fb}$ is infinite on the right axis and $U_{bb}$ is infinite on the upper axis. The filled diamonds present the Mott transition points and the empty squares the mixing-demixing transition points. The solid lines are just the guides for eyes. MI denotes the Mott insulator.}
\end{figure}

Figure 7 shows the snapshots of typical configurations of the Case-1 particles in the mixed Mott insulator (a) and in the demixed Mott insulator (b). The particles fill every site uniformly and make the system insulating both in (a) and (b), but the fermions and bosons are mixed in (a) while separated in (b).
\begin{figure}
\begin{tabular}{cc}
\resizebox{30mm}{!}{\includegraphics{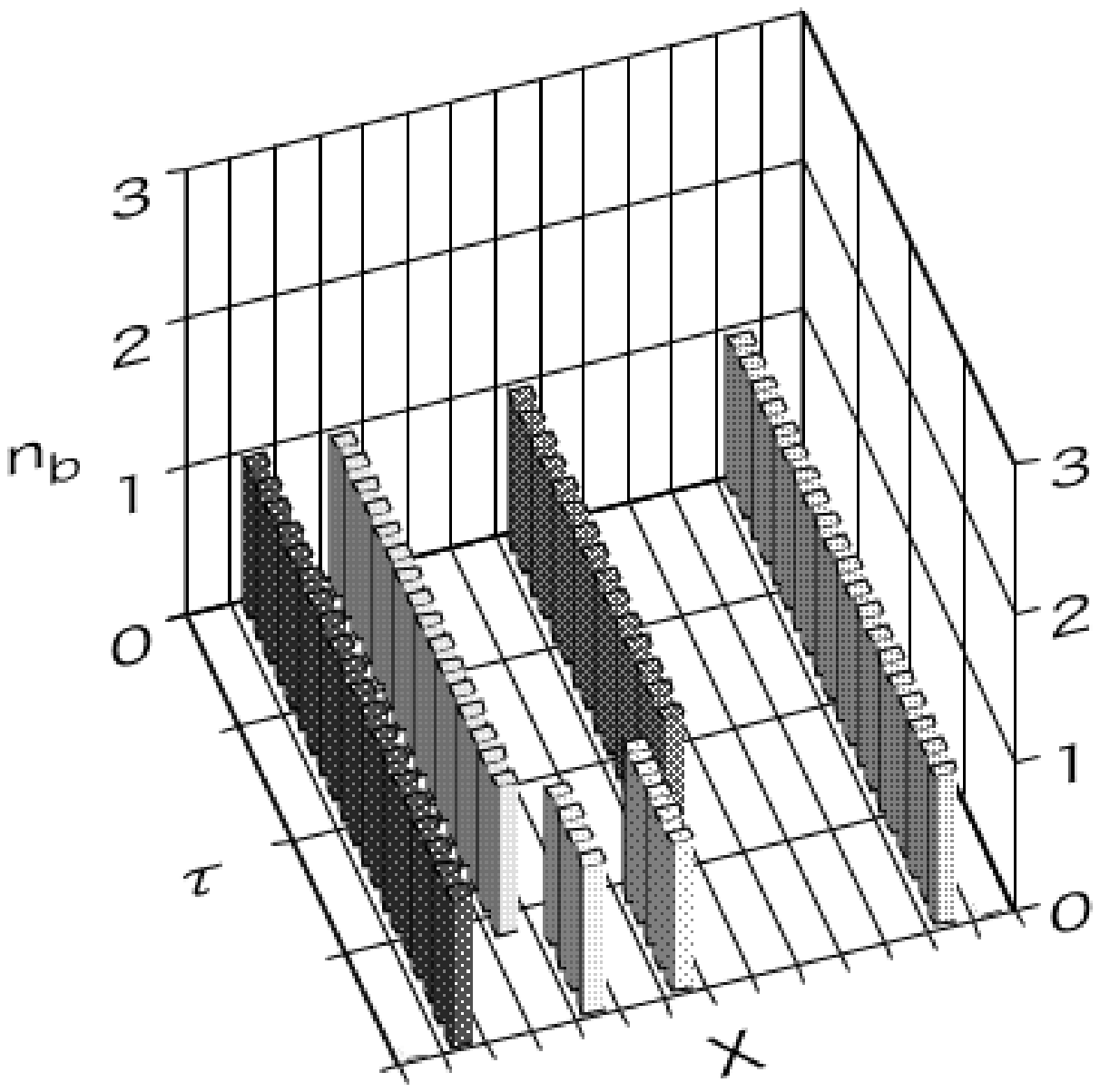}} &\resizebox{30mm}{!}{\includegraphics{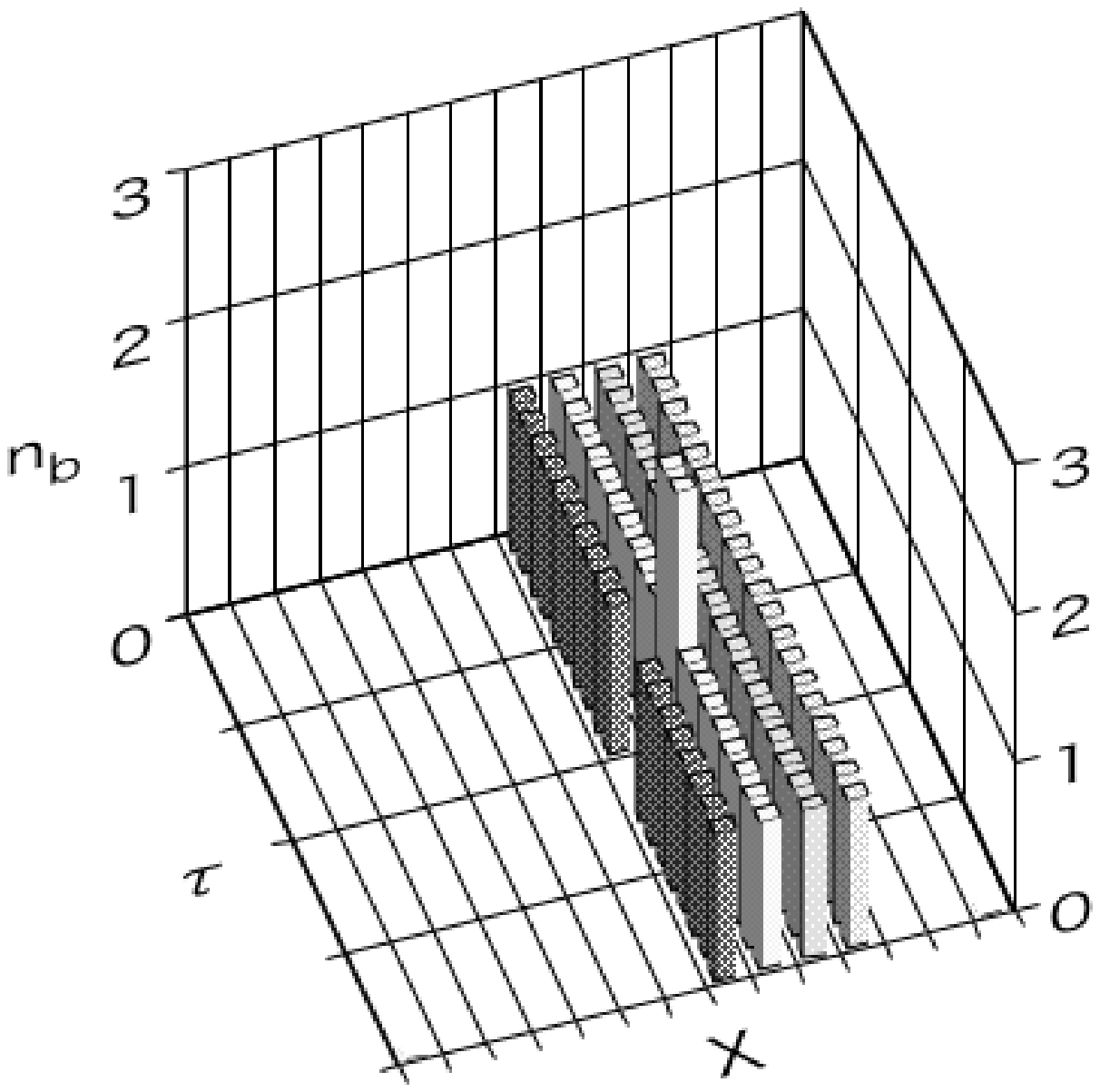}}\\
\resizebox{30mm}{!}{\includegraphics{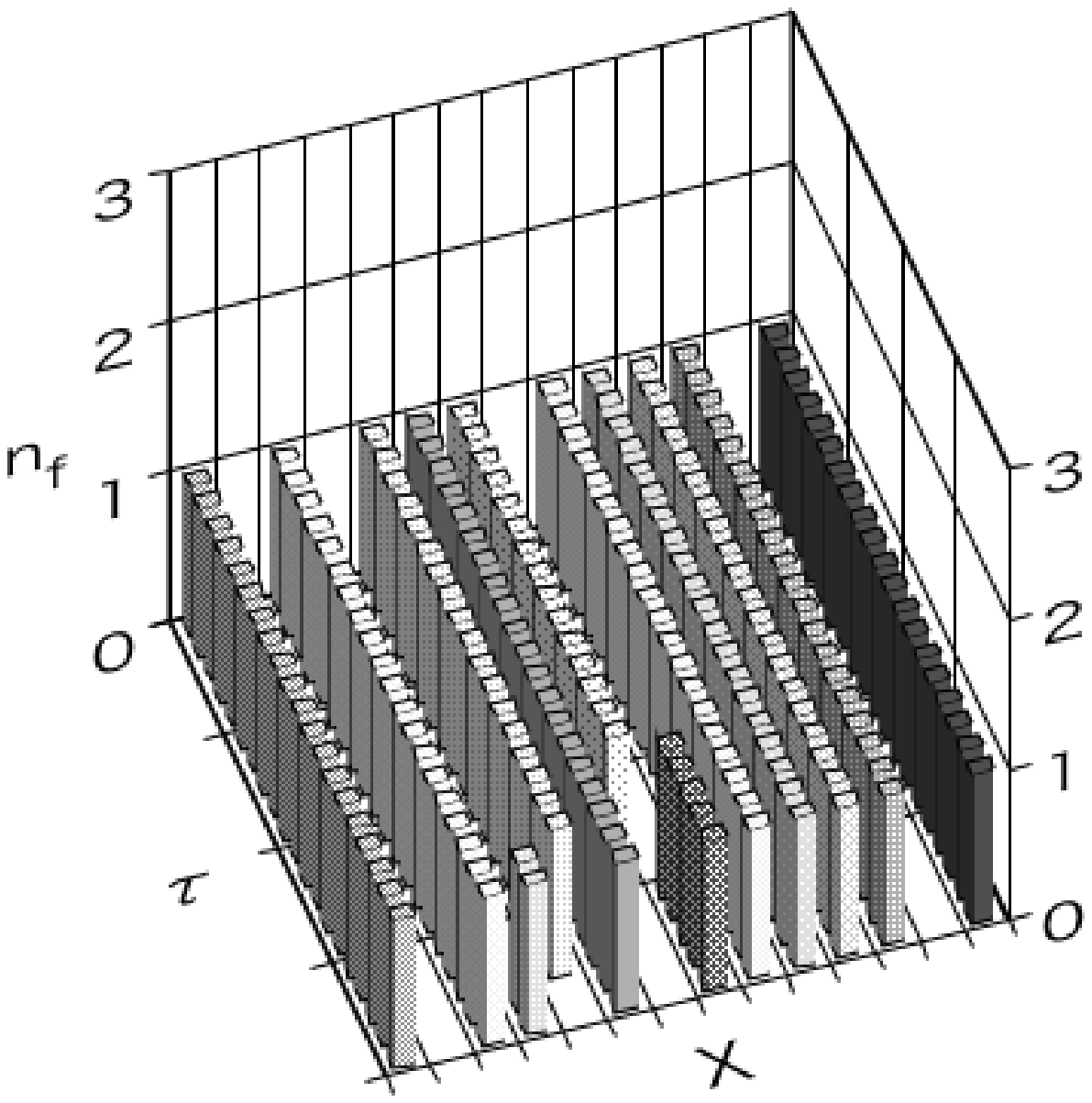}} &\resizebox{30mm}{!}{\includegraphics{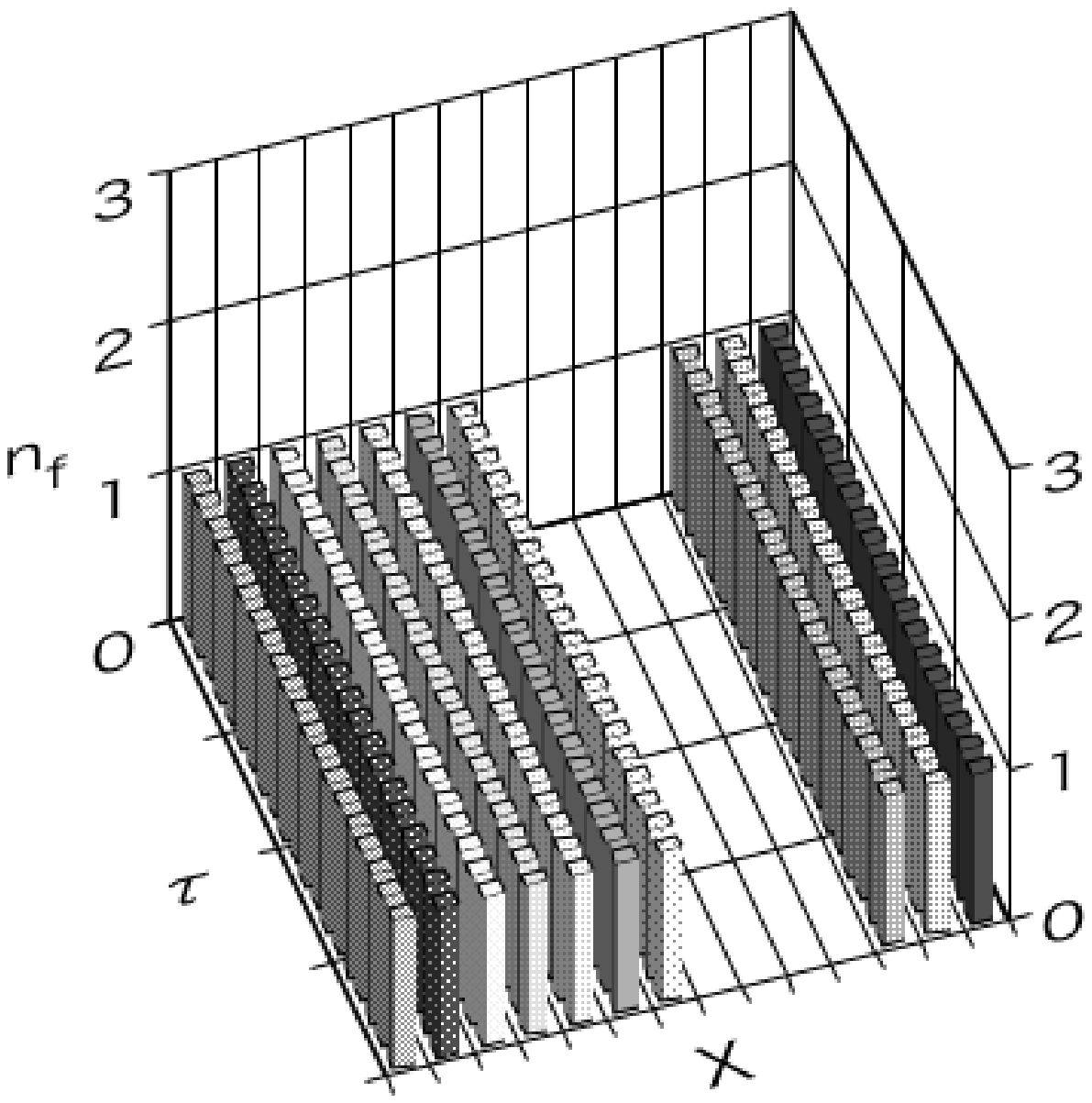}}\\
(a) & (b)
\end{tabular}
\caption{\label{fig7}Snapshots of the particles in the Case 1. (a): The mixed Mott insulator with $U_{fb}=7$ and $U_{bb}=6$. (b): The demixed Mott insulator with $U_{fb}=14$ and $U_{bb}=4$. The upper figure shows the boson configuration and the lower the fermion configuration.}
\end{figure}
\section{Conclusion}
We investigated one-dimensional boson-fermion mixtures with one particle per site. Focuses were placed on two cases. In the Case 1 the fermion filling and boson filling are both incommensurate but the total filling is commensurate, and in the Case 2 the fermion filling and boson filling are both commensurate. The Monte Carlo simulations revealed the occurrence of the mixing-demixing transition and the Mott transition. The system showed the mixing-demixing transition at a certain value of $U_{fb}$ and the transition point shifted to the larger $U_{fb}$ side as $U_{bb}$ increases, which was qualitatively similar to the incommensurate filling case \cite{we}. We found the Mott transition occurred in both mixing and demixing phases at a sufficiently large interaction. The phase diagrams of the Case 1 and 2 were obtained and clearly exhibited the presence of the mixed Mott insulator and the demixed Mott insulator.
\begin{acknowledgments}
The authors would like to thank Prof. Jo and Prof. Oguchi for their continuous support.
\end{acknowledgments}

\end{document}